\RequirePackage{fix-cm}
\documentclass[twocolumn,epjc3]{svjour3}  
\smartqed  
\RequirePackage{graphicx}
\usepackage{color}
\usepackage{textcomp}
\usepackage{gensymb}
\usepackage{cite}  
\journalname{Eur. Phys. J. A}
\begin{document}

\title{Single $\pi^0$ Production Off Neutrons Bound in Deuteron with Linearly Polarized Photons
}


\author{C.~Mullen\thanksref{addr2}
        \and S.~Gardner\thanksref{addr2}
        \and D.~I.~Glazier\thanksref{e1,addr2}
        \and S.~J.~D.~Kay\thanksref{addr4,addr3}
        \and K.~Livingston\thanksref{addr2}
        \and I.~I.~Strakovsky\thanksref{addr5}
        \and R.~L.~Workman\thanksref{addr5}
        \and S.~Abt\thanksref{addr6}
        \and P.~Achenbach\thanksref{addr7}
        \and F.~Afzal\thanksref{addr8}
        \and Z.~Ahmed\thanksref{addr3}
        \and C.~S.~Akondi\thanksref{addr10}
        \and J.~R.~M.~Annand\thanksref{addr2}
        \and M.~Bashkanov\thanksref{addr17}
        \and R.~Beck\thanksref{addr11}
        \and M.~Biroth\thanksref{addr7}
        \and N.~S.~Borisov\thanksref{addr11}
        \and A.~Braghieri\thanksref{addr9}
        \and W.~J.~Briscoe\thanksref{addr5}
        \and F.~Cividini\thanksref{addr7}
        \and C.~Collicott\thanksref{addr7}
        \and S.~Costanza\thanksref{addr9}
        \and A.~Denig\thanksref{addr7}
        \and M.~Dieterle\thanksref{addr6}
        \and E.~J.~Downie\thanksref{addr5}
        \and P.~Drexler\thanksref{addr7}
        \and S.~Fegan\thanksref{addr17}
        \and M.~I.~Ferretti-Bondy\thanksref{addr7}
        \and D.~Ghosal \thanksref{addr6}
        \and I.~Gorodnov\thanksref{addr11}
        \and W.~Gradl\thanksref{addr7}
        \and M.~G\"unther\thanksref{addr6}
        \and G.~Gurevic\thanksref{addr13}
        \and L.~Heijkenskj\"{o}ld\thanksref{addr7}
        \and D.~Hornidge\thanksref{addr16}
        \and G.~M.~Huber\thanksref{addr3}
        \and N.~Jermann\thanksref{addr6}
        \and A.~Kaeser\thanksref{addr6}
        \and M.~Korolija\thanksref{addr18}
        \and V.~L~Kashevarov\thanksref{addr7}
        \and B.~Krusche\thanksref{addr6}
        \and V.~V.~Kulikov\thanksref{addr19}
        \and A.~Lazarev\thanksref{addr8}
        \and S.~Lutterer\thanksref{addr6}
        \and I.~J.~D.~MacGregor\thanksref{addr2}
        \and D.~M.~Manley\thanksref{addr10}
        \and P.~P.~Martel\thanksref{addr7}
        \and M.~A.~Martemianov\thanksref{addr19}
        \and C.~Meier\thanksref{addr6}
        \and R.~Miskimen\thanksref{addr12}
        \and M.~Mocanu\thanksref{addr17}
        \and E.~Mornacchi\thanksref{addr7}
        \and A.~Neganov\thanksref{addr11}
        \and M.~Oberle\thanksref{addr6}
        \and M.~Ostrick\thanksref{addr7}
        \and P.~Otte\thanksref{addr7}
        \and D.~Paudyal\thanksref{addr3}
        \and P.~Pedroni\thanksref{addr9}
        \and A.~Powell\thanksref{addr2}
        \and S.~N.~Prakhov\thanksref{addr7}
        \and G.~Reicherz\thanksref{addr20}
        \and G.~Ron\thanksref{addr14}
        \and T.~Rostomyan \thanksref{addr6}
        \and C.~Sfienti\thanksref{addr7}
        \and V.~Sokhoyan\thanksref{addr7}
        \and K.~Spieker\thanksref{addr8}
        \and O.~Steffen\thanksref{addr7}
        \and Th.~Strub\thanksref{addr6}
        \and I.~Supek\thanksref{addr18}
        \and A.~Thiel\thanksref{addr8}
        \and M.~Thiel\thanksref{addr7}
        \and A.~Thomas\thanksref{addr7}
        \and M.~Unverzagt\thanksref{addr7}
        \and Yu.~A.~Usov\thanksref{addr11}
        \and S.~Wagner\thanksref{addr7}
        \and N.~K.~Walford\thanksref{addr6}
        \and D.~P.~Watts\thanksref{addr17}
        \and D.~Werthm\"uller\thanksref{addr2}
        \and J.~Wettig\thanksref{addr7}
        \and L.~Witthauer\thanksref{addr6}
        \and M.~Wolfes\thanksref{addr7}
        \and N.~Zachariou\thanksref{addr17}
        \\
        \centering{(A2 Collaboration at MAMI)}
}

\thankstext{e1}{e-mail: Derek.Glazier@glasgow.ac.uk}

\institute{
    SUPA, School of Physics and Astronomy, University of Glasgow, Glasgow G12 8QQ, UK 
    \label{addr2}
    \and
    SUPA, School of Physics and Astronomy, University of Edinburgh, Edinburgh EH9 3FD, UK
    \label{addr4}
    \and
    Department of Physics, University of Regina, Regina, SK S4S 0A2, Canada
    \label{addr3}
    \and
    Department of Physics, Institute for Nuclear Studies, The George Washington University, Washington, DC 20052, USA
    \label{addr5}
    \and
    Institut f\"ur Physik, University of Basel, Basel CH-4056, Switzerland
    \label{addr6}
    \and
    Institut f\"ur Kernphysik, Johanes Gutenberg-University Mainz, D-55099 Mainz, Germany
    \label{addr7}
    \and
    Helmholtz--Institut f\"ur Strahlen- und Kernphysik, University of Bonn, Bonn D-53115, Germany
    \label{addr8}
    \and
     INFN Sezione di Pavia, I-27100 Pavia, Italy
    \label{addr9}
    \and
    Kent State University, Kent, OH 44242, USA
    \label{addr10}
    JINR, 141980 Dubna, Russia
    \label{addr11}
    \and
    University of Massachusetts, Amherst, MA 01003, USA
    \label{addr12}
    \and
    Institute for Nuclear Research, 125047 Moscow, Russia
    \label{addr13}
    \and
    Racah Institute of Physics, Hebrew University of Jerusalem, Israel
    \label{addr14}
    \and
    Mount Allison University, Sackville, NB E4L3B5, Canada
    \label{addr16}
    \and
    Department of Physics, University of York, Heslington, York Y010 5DD, UK
    \label{addr17}
    \and
    Rudjer Boskovic Institute, Zagreb HR-10000, Croatia
    \label{addr18}
    \and
    NRC ”Kurchatov Institute” - ITEP, Moscow 117218, Russia
    \label{addr19}
    \and
    Institut f\"ur Experimentalphysik, Ruhr-University of Bochum, Bochum D-44801, Germany
    \label{addr20}
\\
}

\date{Received: date / Accepted: date}

             
\maketitle

\begin{abstract}
The quasifree $\overrightarrow{\gamma} d\to\pi^0n(p)$ photon beam asymmetry, $\Sigma$, has been measured at photon energies, $E_\gamma$, from 390 to 610~MeV, corresponding to center of mass energy from 1.271 to 1.424~GeV, for the first time. The data were collected in the A2 hall of the MAMI electron beam facility with the Crystal Ball and TAPS calorimeters covering pion center-of-mass angles from 49 to 148$\degree$. In this kinematic region, polarization observables are sensitive to contributions from the $\Delta (1232)$ and $N(1440)$ resonances. The extracted values of $\Sigma$ have been compared to predictions based on partial-wave analyses (PWAs) of the existing pion photoproduction database. Our comparison includes the SAID, MAID, and Bonn-Gatchina analyses; while a revised SAID fit, including the new $\Sigma$ measurements, has also been performed. In addition, isospin symmetry is examined as a way to predict $\pi^0n$ photoproduction observables, based on fits to published data in the channels $\pi^0p$, $\pi^+n$, and $\pi^-p$. 

\end{abstract}

\noindent

\section{Introduction}
\label{Sec:Intro}

Knowledge of the $N^\ast$ and $\Delta^\ast$ resonance decay couplings to nucleons and photons is largely restricted to charged states. Increasing the body of neutron-target measurements will allow a more highly constrained study of neutral states and their $n\gamma$ couplings. The four charge channels ($\pi^0p$, $\pi^+n$, $\pi^-p$, and $\pi^0n$) of pion photoproduction can be described in terms of three isospin amplitudes. This gives the possibility of predicting properties of one channel based on sufficiently detailed measurements of the other three. The $\pi^0n$ channel is the least-studied and was the subject of this experiment.

Most existing $\gamma n$ data are unpolarized and provide around 1900 $\pi^0n$ photoproduction data points, spanning the full nucleon resonance region~\cite{Ireland:2019uwn} (Table~\ref{tab:tbl1}). The $\overrightarrow{\gamma} n\to\pi^0n$ beam asymmetry, $\Sigma$, was previously measured by the GRAAL Collaboration~\cite{DiSalvo:2009zz}. The beam asymmetry measures the relative strength of the production with respect to the plane of photon linear polarisation. Their measurements covered beam energy, $E_\gamma$, from 703 to 1475~MeV, corresponding to a centre-of-mass energy, $W$, range from 1.484 to 1.912~GeV, just above the current results. 

Recently, the A2 Collaboration at MAMI published high-quality unpolarized measurements for $\pi^0$ photoproduction off a neutron below $E_\gamma$ = 813~MeV~\cite{Briscoe:2019cyo}. The present data extend the range of the previous GRAAL polarized measurements~\cite{DiSalvo:2009zz}, for $\pi^0n$ photoproduction. Further A2 Collaboration measurements of the $\pi^0n$ $E$ asymmetry, with longitudinal polarized target and circularly polarized photons, for $E_\gamma$ = 216 -- 1606~MeV~\cite{Costanza:2020wnx}, extend previous A2 $\pi^0n$ $E$ measurements~\cite{Dieterle:2017myg}. These data will provide the basis for better-constrained $\gamma n$ decay amplitudes in the near future. 

Apart from lower-energy inverse reaction $\pi^-p\to\gamma n$ measurements, the extraction of the two-body $\gamma n\to\pi^-p$ and $\gamma n\to\pi^0n$ observables requires the use of a model-dependent nuclear correction, which mainly comes from final-state interaction (FSI) effects. In several papers, the GWU-ITEP group have shown that the FSI corrections on unpolarized cross sections are less than 20\% (see, for instance, Refs.~\cite{Tarasov:2011ec,Mattione:2017fxc,Tarasov:2015sta}). As polarization asymmetries measure ratios of cross sections, FSI effects are expected to have a considerably smaller effect on these, including $\Sigma$, and will be comparable, or less than our quoted systematic uncertainties from experimental sources. In this publication, $\Sigma$ for the neutron bound in a deuteron is presented uncorrected for potential FSI effects so as not to add any model dependence to the results.
\begin{table*}[htb!]

\centering \protect\caption{Published data for $\gamma n\to\pi^0n$ reaction as given in the SAID database~\protect\cite{Briscoe:2020zzz}:
    1st column is the observable, 
    2nd column is the number of energy bins,
    3rd column is the number of data points.}
    
\vspace{2mm}
{%
\begin{tabular}{|ccccccccc|}
\hline
Observable        & Nexp & Ndata & E$_\gamma$(min) & E$_\gamma$(max) & $\theta$(min) & $\theta$(max) & Laboratory  & Ref \tabularnewline
                  &      &       &    (MeV)        &      (MeV)      &   ($\degree$)  & ($\degree$) 
                  &          &   \tabularnewline
\hline
$d\sigma/d\Omega$ & 9    & 9     & 208             & 373             & 147           & 148           & MAMI     &  \protect\cite{Kossert:2003zf}\tabularnewline 
                  & 27   & 492   & 290             & 813             & 18            & 162           & MAMI     & 
\protect\cite{Briscoe:2019cyo}\tabularnewline 
                  & 40   & 43    & 299             & 889             & 70            & 130           & Tokyo    & 
\protect\cite{Ando:1977aa}\tabularnewline
                  & 49   & 931   & 446             & 1427            & 32            & 162           & MAMI     &
\protect\cite{Dieterle:2018adj}\tabularnewline
                  & 42   & 42    & 455             & 905             & 45            & 143           & Tokyo    &
\protect\cite{Hemmi:1973ii}\tabularnewline 
                  & 35   & 35    & 462             & 784             & 60            & 135           & Frascati &
\protect\cite{Bacci:1972bh}\tabularnewline
                  & 3    & 28    & 911             & 1390            & 3             & 91            & SLAC     &
\protect\cite{Clinesmith:1967zn}\tabularnewline
\hline
  $\Sigma$        & 27   & 216   & 703             & 1475            & 53            & 164           & GRAAL                & 
\protect\cite{DiSalvo:2009zz}\tabularnewline
\hline
  $E$             & 17   & 151   & 446             & 1427            & 46            & 154           & MAMI                 & 
\protect\cite{Dieterle:2017myg}\tabularnewline 
\hline
\end{tabular}} \label{tab:tbl1}
\end{table*}

The organization for this paper is as follows. In 
Section~\ref{Sec:Experiment}, details of the A2 experiment and detectors are given; Section~\ref{Sec:DataAnalysis} outlines the event selection; 
Section~\ref{Sec:Background}  reviews the background subtraction;
Section~\ref{Sec:Normalization} covers the determination of the photon asymmetry; and
Section~\ref{Sec:Syst} outlines the dominant sources of systematic uncertainty;
Section~\ref{Sec:PWA} outlines the PWA methods used in the fits and predictions compared to data. Finally, 
Section~\ref{Sec:Results} presents the results and interpretation of the present A2 $\Sigma$ data. 

\section{Experiment}
\label{Sec:Experiment}

The reaction $\overrightarrow{\gamma}$d $\rightarrow \pi^0$n(p) was measured at the Mainzer Microtron (MAMI) electron accelerator facility, in August 2016. The 1.5~GeV MAMI electron beam, incident on an aligned diamond radiator, produced a photon beam via coherent bremsstrahlung, with significant linear polarization up to photon energies of $610$~MeV. The energy of the photon beam was measured using the Glasgow-Mainz Tagged Photon Spectrometer with a resolution of around $4$~MeV. This spectrometer measured the position of the degraded post-bremsstrahlung electron on a plastic scintillator focal plane consisting of 353 elements after traversing a 1.8~T magnetic dipole field~\cite{McGeorge:2007tg}. The energy of the detected electron, and therefore also the energy of the photon, was deduced from this position. 

\begin{figure}
\begin{center}
\includegraphics[height=2.3in, keepaspectratio]{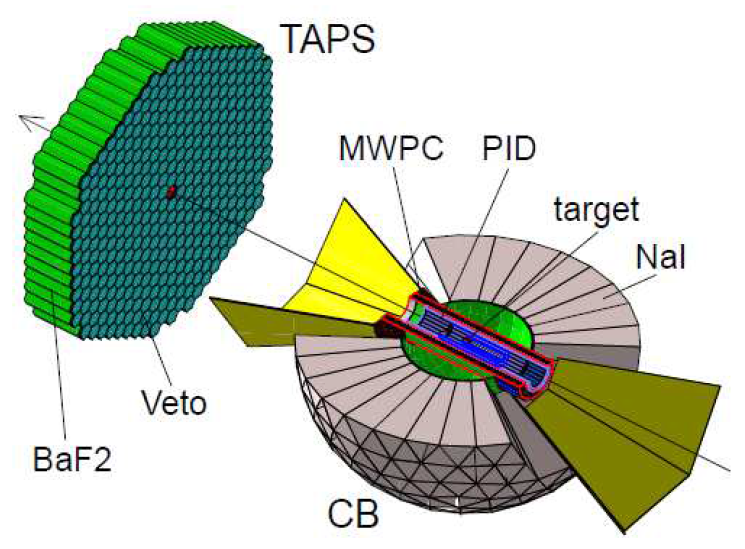}
\end{center}

\vspace{-1mm}
        \protect\caption{Set-up of the A2 experiment. 
        CB shows a NaI(Tl) calorimeter,
        TAPS shows a BaF$_2$ and PbWO$_4$ calorimeter, 
        PID shows a plastic scintillator detector for particle identification, 
        MWPC are two cylindrical multiwire proportional chambers, and
        target shos the liquid deuterium target $LD_2$ (see text for details).}
        \label{fig:setup}
\end{figure}

The photon beam interacted in a 10~cm long liquid deuterium target ($LD_2$). The reaction products were detected in two calorimeters: the Crystal Ball (CB), a highly segmented array of 672 NaI(Tl) crystals arranged in a sphere centered on the target cell~\cite{Starostin:2001zz}; and the TAPS calorimeter, a forward wall of 366 BaF$_2$ and 72 PbWO$_4$ crystals arranged 1.5~m downstream from the CB center~\cite{Gabler:1994ay} (Fig.~\ref{fig:setup}). The CB covers lab. frame angles 21$\degree<\theta_{L}<159\degree$ and TAPS approximately $2\degree<\theta_{L}<20\degree$. The $LD_2$ target cell was surrounded by a Particle Identification Detector (PID) consisting of 24, 30~cm long plastic scintillators arranged in a cylindrical formation. This allowed separation of reactions with a scattered neutron from those with a proton. A Multi-Wire Proportional Chamber barrel (MWPC) provided tracking information for charged particles which were not used for the all-neutral final state investigated here, however here they provided additional proton rejection. Charge particle identification was provided in the case of the TAPS detector by a thin plastic veto layer in front of each crystal. In addition, a 2.6~cm thick graphite cylinder was situated between the PID and the MWPC to be used as the analysing material for a nucleon polarimeter~\cite{Bashkanov:2019mpj}, and was not required for this analysis~\cite{Mullen:2020fgd}.

All simulations used in this analysis were performed with a full detector model using the  Geant4 \cite{GEANT4} toolkit.
\vspace{7mm}

The linear polarization of the photons was produced by coherent bremsstrahlung~\cite{Lohmann:1994vz,Timm:1969mf}, with the electron beam scattered coherently from an aligned radiator. A thin diamond crystal (30~$\mu$m) with low mosaic-structure was used to minimize the energy smearing of the coherent spectrum arising from electron multiple scattering effects and crystal defects in the lattice~\cite{Kellie:2005wy}. The alignment of the diamond was carried out using the Stonehenge technique~\cite{Livingston:2008hv} with two orthogonal polarization plane orientations chosen to be at azimuthal angles of $\pm45\degree$ with respect to the equatorial plane of the CB detector. To increase the degree of linear polarization a 2~mm diameter Pb-collimator was installed 2.5~m downstream of the radiator, enhancing the ratio of coherent to incoherently scattered photons that reached the $LD_2$. The degree of linear polarization was determined by calibrating against the linearly polarized photon beam asymmetry for $\pi^0$ production off the proton. This was measured for each photon energy bin and compared to a recent SAID PWA solution including recent high statistics measurements in the same energy range~\cite{Gardner:2016irh}. The ratio of the measured asymmetry to the SAID values gave the photon polarization for each energy bin. The analysis of the proton asymmetry was performed in the same manner as the neutron asymmetry described here. The resulting photon polarization ranged from $15 \%$ at $E_\gamma$ = 390~MeV to a maximum of $55 \%$ at 610~MeV~\cite{Mullen:2020fgd}.

\section{Data Analysis}
\label{Sec:DataAnalysis}

The photon asymmetry, $\Sigma$, for the reaction $\overrightarrow{\gamma}$d $\rightarrow \pi^0$ n(p) has been measured for beam energies in the range 390 -- 610~MeV and a center-of-mass (c.m.) production angle, $\theta$, range of 49$\degree$ -- 148$\degree$. The semi-inclusive final state of interest included the recoiling neutron and $\pi^0$, and omitted the spectator proton. It was identified by detecting three neutral particles, two $\gamma$s stemming from the decay of the $\pi^0$, and a neutron as the third. The energy of the three particles was measured by the CB and TAPS calorimeters which, in coincidence with a tagged photon, allowed the reaction to be reconstructed. The classification of a neutral state was made if there were no hits in the PID or MWPC detectors for the CB or the TAPS veto layer. The spectator proton was not considered as it typically did not have sufficient energy to reach the calorimeters. The $\pi^0$ was reconstructed from the combination of two of the three particles detected. All combinations were considered, and any incorrect 2$\gamma$ combinations were removed, either by subsequent cuts or by background subtraction.
\begin{table}[ht] 

        \protect\caption[sPlots Cuts]{A summary of the loose cuts applied to the data 
        before the sPlots fits.}
\centering
\begin{tabular}{|l|l|l|}
\hline
\textbf{Variable}  & \textbf{Cut Range}                         & \textbf{Units} \\ \hline
Tagged Time        & $-80$ \textless{} t$_{\pi^{0}}$  \textless{} 20   & ns        \\ \hline
Coplanarity        & -50 \textless{} $\Delta\phi$   \textless{} 50   & degree    \\ \hline
Missing Mass       & 1850 \textless{} M$_\mathrm{miss}$     \textless{} 2300 & MeV/c$^2$ \\ \hline
Cone Angle         & 0   \textless{} $\theta_\mathrm{Cone}$ \textless{} 0.5  & radian    \\ \hline
Invariant Mass     & 80  \textless{} M$_\mathrm{inv}$      \textless{} 200  & MeV/c$^2$ \\ \hline
Spectator Momentum & 0   \textless{} $\left| \mathrm{P}_\mathrm{spec} \right|$ \textless{} 200  & MeV/$c$   \\ \hline
\end{tabular}
        \label{Table:LooseCuts}
\end{table}	

Preliminary cuts were placed on a number of variables. These were guided by simulated signal and background channels to ensure no actual signal was lost and are given in Table~\ref{Table:LooseCuts}. The missing mass was constructed using the mass of the missing 4-momentum defined as:
\begin{equation}
    P_\mathrm{miss} = P_\mathrm{beam} + P_\mathrm{d} - P_{\pi^0},
    \label{eq:eq1}
\end{equation}
where P$_\mathrm{beam}$ is the 4-momentum for the beam photon, P$_\mathrm{d}$ is the 4-momentum of the stationary deuterium target, and P$_{\pi^0}$ is the 4-momentum of the detected $\pi^0$. The mass of this missing 4-momentum gave a peak distributed around the summed mass of the two nucleons with some extra smearing from initial Fermi motion within the deuteron. When this initial momentum was low, as was generally the case, the resulting spectator proton momentum was also low and the reaction was approximately two-body with the $\pi^0$ and participant nucleon being nearly coplanar in $\phi$:
\begin{equation}
    \Delta\phi = \phi_{\pi^0} - \phi_{n - 180\degree}~ \sim 0\degree,
    \label{eq:eq2}
\end{equation}
where $\phi_{\pi^0}$ is the reconstructed azimuthal angle of the $\pi^0$ and $\phi_{n - 180\degree}$ the azimuthal angle of the detected neutron after rotation by $180\degree$ around the z-axis.

The detection of the $\pi^0$ and the neutron allowed the construction of the difference between the detected nucleon polar angle and the nucleon polar angle reconstructed from the $\pi^0$ assuming a stationary initial state neutron.  This gives the definition of the ``Cone Angle'', $\theta_\mathrm{Cone}$.

The momentum of the spectator proton was also used to distinguish the quasi-free final state. It was calculated via:
\begin{equation}
    P_\mathrm{spec} = P_\mathrm{beam} + P_\mathrm{d} - P_{\pi^0} - P_\mathrm{n},
    \label{eq:eq3}
\end{equation}
where $P_\mathrm{n}$ is the 4-momentum of the detected participant nucleon. The magnitude of momentum for the participant nucleon was calculated using conservation of momentum and energy in the three-body final state, using the measured $\pi^0$ momentum and the neutron direction, as given by its cluster hit position in the CB.

\section{Background Subtractions}
\label{Sec:Background}

To further isolate the true $\pi^0$n final state, several sources of background had to be subtracted from the selected events. The main sources of these backgrounds were: random electrons in the photon tagger; background to the two photon combination giving the $\pi^0$; and other reactions producing the same detected particles as the $\pi^0$n(p) reaction. The sPlot technique was used to remove these background events using a separate discriminatory variable for each source to produce event-by-event weights termed sWeights, for full details see ~\cite{Pivk:2004ty}. The sWeights are normalized using the relative yields and covariance matrix of the signal and background derived from the fits. Weights corresponding to regions of high background are negative and effectively subtract off this contribution to the distribution. This is similar to how a ``sideband subtraction'' method works, but is more generally applicable. Consecutive fits were performed applying the sWeights from the previous fit. An sPlot fit to the discriminatory variables using appropriate probability density functions (PDFs) derived from simulated event samples, determined the yields of the different event species. The covariance matrix of this yield fit was then used to calculate the sWeights associated with each event in the fitted sample. Including these sWeights in the subsequent observable fits allowed determination of the photon asymmetry for our signal. The fits to the discriminatory variables are described in the following sections.

\subsection{Random Tagged Photons}

\begin{figure}
\centerline{\includegraphics[height=6cm]{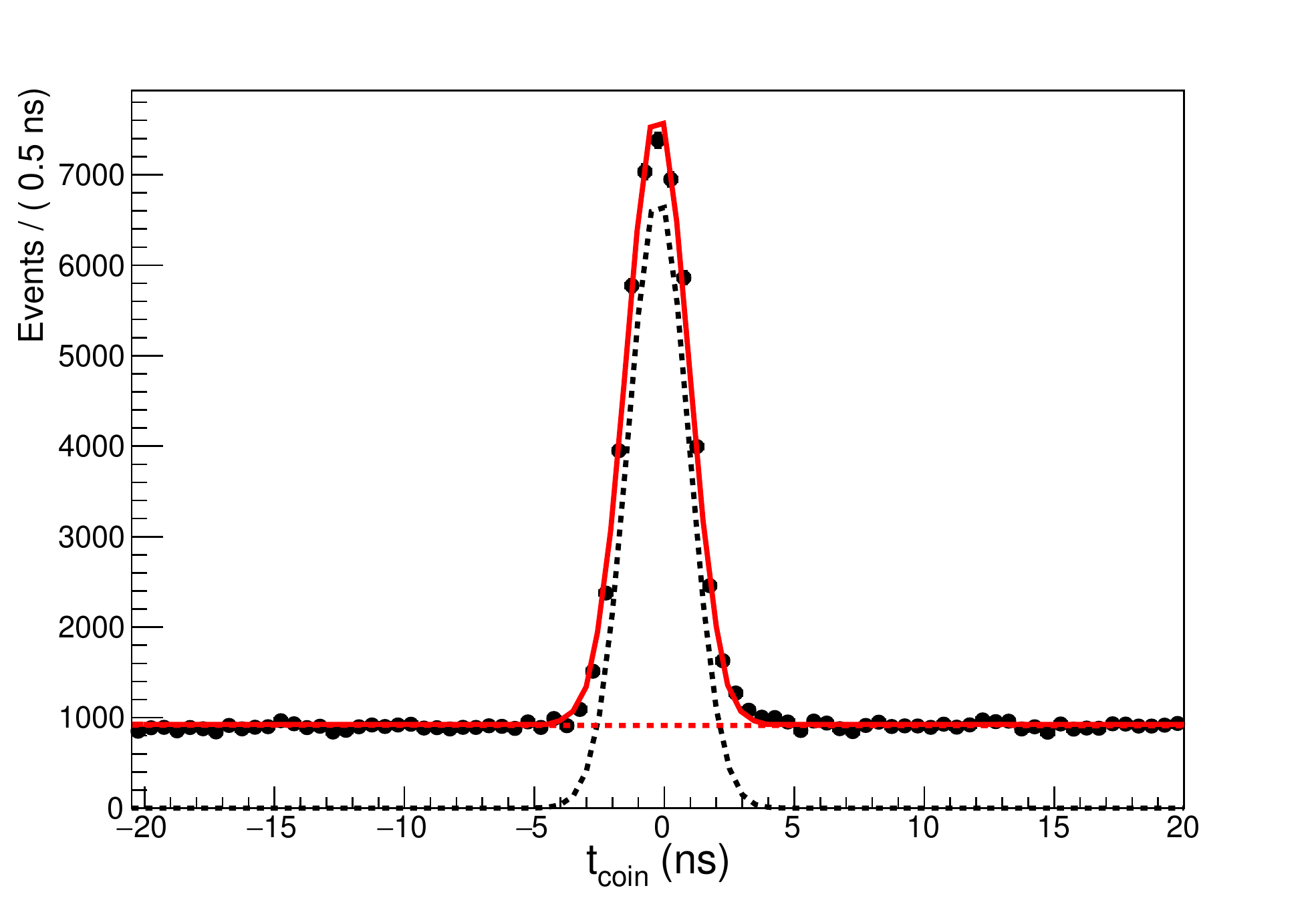}}

        \protect\caption{The timing coincidence spectra between the photon beam tagger 
        and the calorimeters. Black points are data; red solid line is full fit result; 
        dashed black is signal Gaussian function; dashed red is flat background function. 
        This fit was used to produce weights to subtract the random background events.}
        \label{fig:tcoinc} 
\end{figure}

Random coincidences with background electrons in the photon tagger were removed via the coincidence time between the  $\pi^0$ and the tagged beam photon as given by:
\begin{equation}
    t_\mathrm{coin} = \frac{t_{\gamma1} + t_{\gamma2}}{2} - t_\mathrm{tagger},
    \label{eq:eq4}
\end{equation}
with the time of the electron in the tagger, $t_\mathrm{tagger}$ and the time in the calorimeters of the 2 photons $t_{\gamma1,2}$. This resulted in a timing distribution strongly peaked at zero with a flat random background, as shown in Fig.~\ref{fig:tcoinc} for the bin $E_\gamma = 610$~MeV and $\cos\theta_{CM} = 0.05$. In this case, a Gaussian PDF was used for the signal with a uniform background function.

\subsection{Background in $2\gamma$ Invariant Mass}

\begin{figure}
\centerline{\includegraphics[height=6cm]{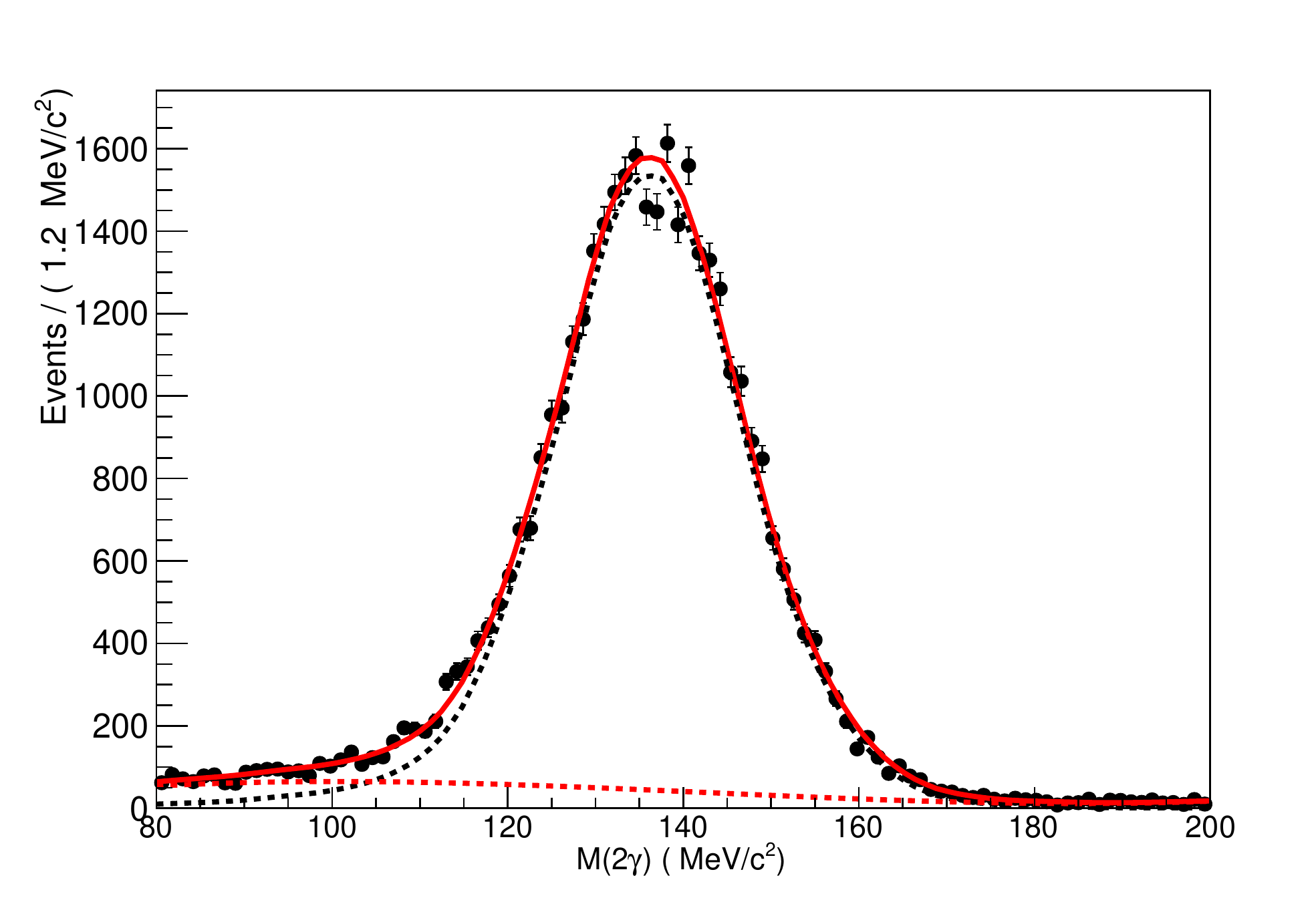}}

        \protect\caption{The invariant mass of the two detected $\gamma$s. Black 
        points are data; red solid line is full fit result; dashed black is 
        simulated signal function; dashed red is third degree Chebychev polynomial. 
        This fit was used to produce weights to subtract background events that did not have 
        a $\pi^0$. }
        \label{fig:pi0mass}
\end{figure}

Background to real $\pi^{0}$ decays in the two $\gamma$ invariant mass distribution can arise from a wrong combination of the three neutral clusters, or multiple clusters created by one actual particle. These background sources will not give a peaking structure in the invariant mass distribution and were thus subtracted using the sPlot technique. The $\pi^{0}$ signal PDF was taken from a histogram template of simulated events, while the background was modelled by a third order Chebychev polynomial. An example fit is shown for the bin $E_\gamma = 610$~MeV and $\cos\theta = 0.05$ in Fig.~\ref{fig:pi0mass}.  In this mass range, the $\pi^0$ signal was typically around $90\%$ of the total events.

\subsection{Background to the $\pi^0$n(p) final state}

\begin{figure}
\centerline{\includegraphics[height=6cm]{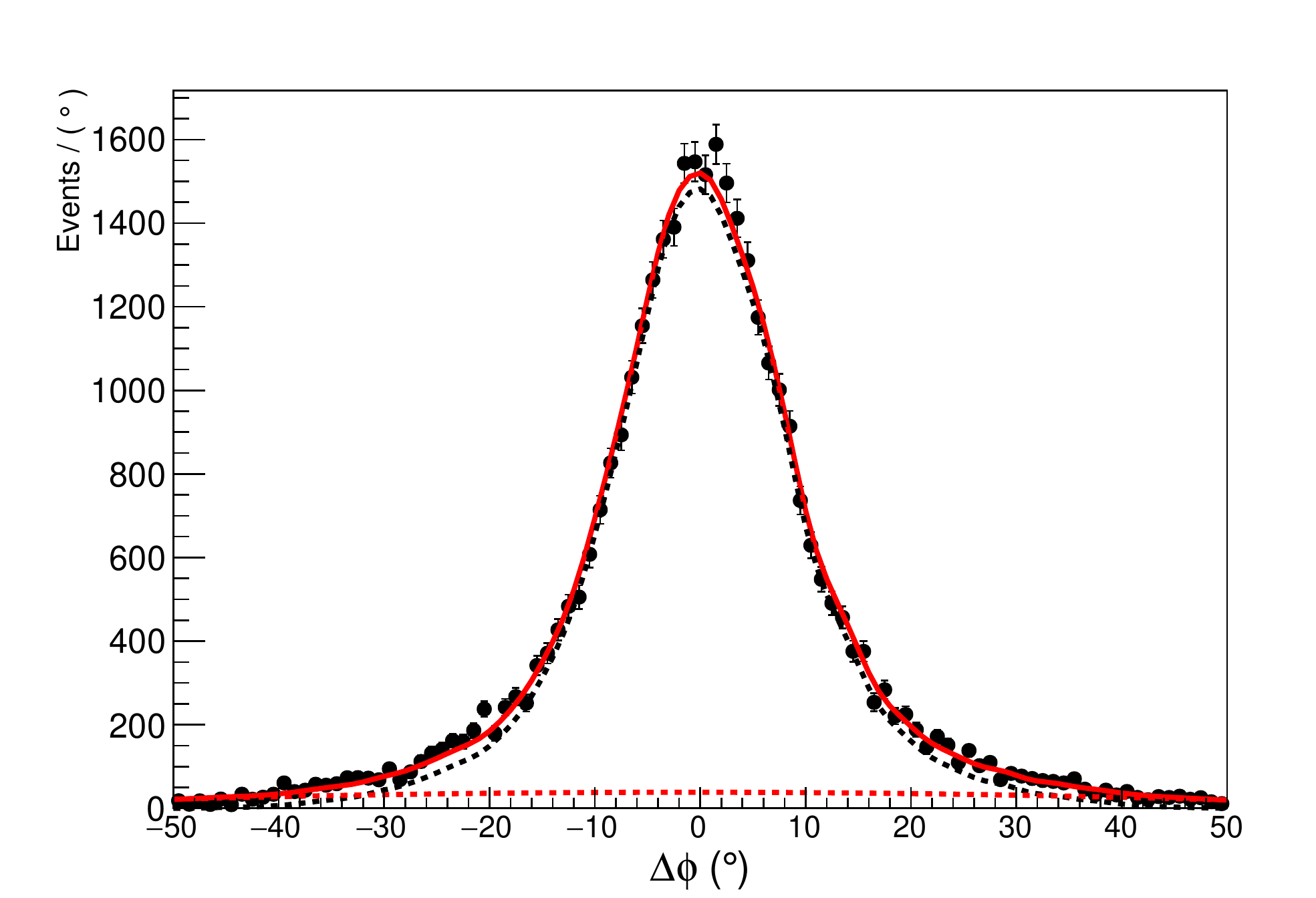}}

        \protect\caption{The coplanarity between the $\pi^0$ and detected neutron. Black points are data; red solid line is full fit result; dashed black is simulated signal function; dashed red is second degree Chebychev polynomial. This fit was used to produce weights to subtract events that did not originate from the $\overrightarrow{\gamma}$d $\rightarrow \pi^0$ n(p) final state.}
        \label{fig:coplane}
\end{figure}

Background to the final state may come, for example, from events in which more than one pion is produced. This background is reduced with the loose cuts given in Table~\ref{Table:LooseCuts}. To determine the sWeights for subtracting the residual background, the coplanarity given in Eq.~\ref{eq:eq2} was used. The signal PDF shape was given by simulated data and the background by a second order Chebychev polynomial. The resulting fit, for the bin $E_\gamma = 610$~MeV and $\cos\theta = 0.05$, is shown in Fig.~\ref{fig:coplane}.
\begin{figure}
\centerline{\includegraphics[height=5.5cm]{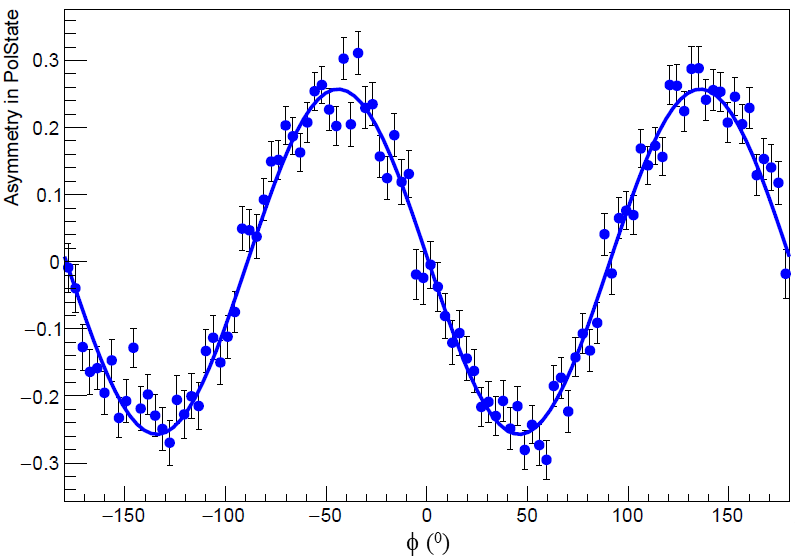}}

        \protect\caption{ Example of fitted asymmetry used to extract $\Sigma$. 
        Blue points are the data and blue solid line is the result of the maximum 
        likelihood fit.}
        \label{fig:fig6}
\end{figure}

%
\section{Determination of the Photon Asymmetry}
\label{Sec:Normalization}

The photon asymmetry, $\Sigma$, quantifies the effect of the polarization of the beam on the excitation of the neutron and its subsequent decay to a pion and nucleon. With a linearly polarized photon beam the differential cross section is
\begin{equation}
    \frac{d\sigma}{d\Omega} = \left(\frac{d\sigma}{d\Omega}\right)_{0} ,
    (1 + P_L \Sigma \cos2\phi ),
    \label{eq:Sigma}
\end{equation}
where $\phi$ is the azimuthal angle of the meson production plane relative to the plane of linear polarization and $P_L$ is the degree of linear polarization. Rotating the orientation of the diamond radiator allowed the plane of linear polarization to flip between $-\frac{\pi}{4}$ and $\frac{\pi}{4}$. Shifting the plane by $\frac{\pi}{2}$ effectively flips the sign of the asymmetry giving two polarization states $P_S = \pm1$.

To extract $\Sigma$ from the measured $\phi$ distributions unbinned extended maximum likelihood fits were performed. The fit function was given by
\begin{equation}
    F(\Sigma:\phi,P_S,P_L) = 1 + \Sigma P_SP_L \cos(2\phi + \phi_0),
    \label{eq:FPDF}
\end{equation}
with $\phi_0$ determined from fits to be $95.8\degree$.

The negative log likelihood function given by
\begin{equation}
    - \ln{L} = -\sum^{N}_iw_i\ln{F(\Sigma:\phi_i,P_{S,i},P_{L,i})} + B(\Sigma),
    \label{eq:Likelihood}
\end{equation}
was minimized using Minuit as part of the ROOFIT~\cite{Verkerke:2003ir} library. Here, $N$ is the number of data events in the $E_\gamma$ and $\theta$ bin, while subscript $i$ refers to the value of the variables for a given event. In particular, $w_i$ represents the value of the sWeight used to subtract background events from the likelihood summation.

The PDF normalization integral term $B(\Sigma)$ was determined by Monte-Carlo integration using simulated data. For this, $P_S$ and $P_L$ values were randomly chosen to match the fluxes and polarization degrees of the real data. This effectively corrected for second order systematic effects due to differences in polarization state luminosity and degree of polarization.

\begin{figure*}
\centerline{\includegraphics[height=1.1\textwidth,angle=90]{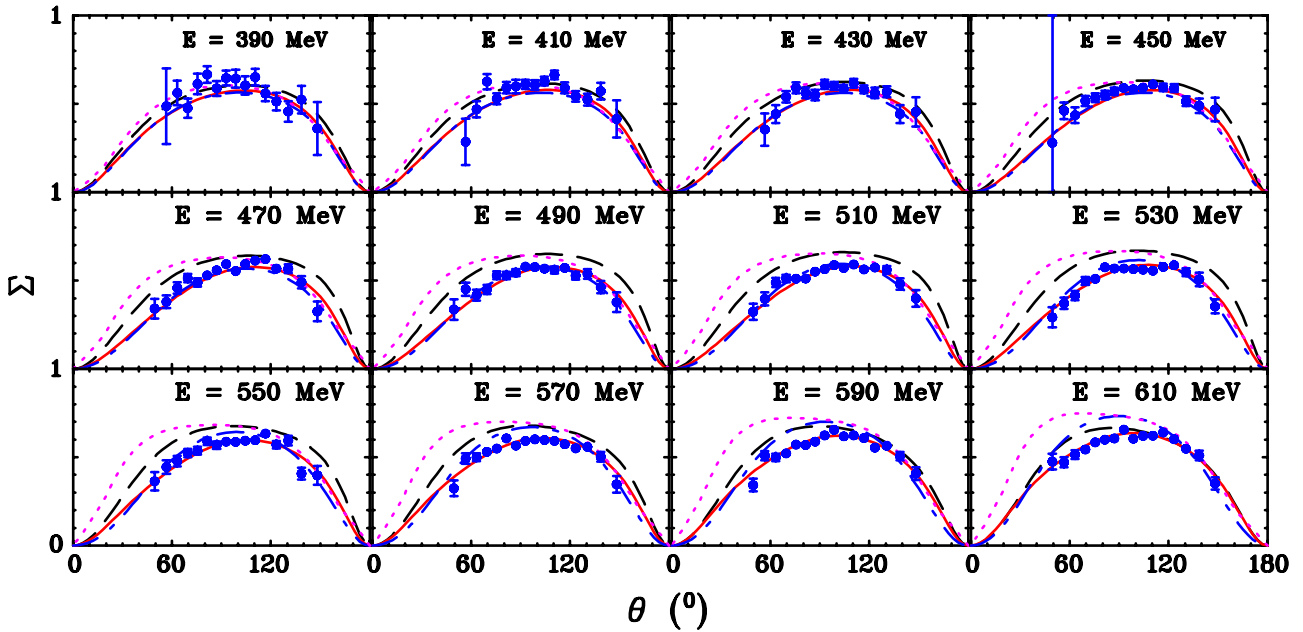}}

\vspace{-50mm}
        \protect\caption{$\Sigma$ for $\overrightarrow{\gamma} n\to \pi^0n$ vs. pion 
        production angle $\theta$ in c.m. frame: A2 (blue filled circles); fit: SAID 
        MU22 (red solid curves), SAID MA19~\protect\cite{Briscoe:2019cyo} (blue 
        dash-dotted curves), Bonn-Gatchina BG2014-02~\protect\cite{Gutz:2014wit} 
        (magenta dotted curves), and MAID2007~\protect\cite{Drechsel:2007if} (black 
        dashed curves). Only angle-dependent uncertainties are shown for all data. Each plot corresponds to a 20 MeV wide bin in E$_\gamma$ with the central value given in each.}
        \label{fig:sigma}       
\end{figure*}

An example of the resulting fitted asymmetry in polarization state compared to the background subtracted data asymmetry, where the plotted asymmetry is calculated as,
\begin{equation}
    A(\phi) = \frac{F(\phi:P_{S}=+1) - F(\phi:P_{S}=-1)}{F(\phi,P_{S}=+1) + F(\phi,P_{S}=-1)}
\end{equation}
is shown in Fig.~\ref{fig:fig6}, for the bin $E_\gamma = 610$~MeV and $\cos\theta = 0.05$, as an illustration. The fits were performed in bins of 20~MeV for $E_\gamma$ and 0.1 in $\cos\theta$. Results are shown in Fig.~\ref{fig:sigma} alongside solutions of various PWAs described in Section~\ref{Sec:PWA}. 

\section{Systematic Uncertainties}
\label{Sec:Syst}

The dominant sources of systematic uncertainty in the results came from the linear polarization calibration and the background subtraction. The fractional difference between a simple cuts-based analysis and the sPlot background subtraction methods gave an estimate of the systematic uncertainty in background subtraction method for each point.
For the cuts-based analysis a similar procedure was performed but, rather than use a sPlot subtraction, tighter cuts were placed on the discriminatory variables to identify a cleaner sample of $n\pi^{0}$ events with residual backgrounds estimated by simulations to be around 3.6$\%$. These cuts are summarised in Table \ref{Table:TightCuts}.
\begin{table}[ht] 

        \protect\caption[Tight Cuts]{A summary of the tight cuts applied to the data to produce a low background event sample.}
\centering
\begin{tabular}{|l|l|l|}
\hline
\textbf{Variable}  & \textbf{Cut Range}                         & \textbf{Units} \\ \hline
Tagged Time        & -5 \textless{} t$_{\pi^{0}}$  \textless{} 5   & ns        \\ \hline
Coplanarity        & -30 \textless{} $\Delta\phi$   \textless{} 30   & degree    \\ \hline
Missing Mass       & 1850 \textless{} M$_\mathrm{miss}$     \textless{} 2100 & MeV/c$^2$ \\ \hline
Cone Angle         & 0   \textless{} $\theta_\mathrm{Cone}$ \textless{} 0.3  & radian    \\ \hline
Invariant Mass     & 110  \textless{} M$_\mathrm{inv}$      \textless{} 160  & MeV/c$^2$ \\ \hline
Spectator Momentum & 0   \textless{} $\left| \mathrm{P}_\mathrm{spec} \right|$ \textless{} 200  & MeV/$c$   \\ \hline
\end{tabular}
        \label{Table:TightCuts}
\end{table}	
While a cuts-only analysis retains some small amount of background, as illustrated in Figs.~\ref{fig:pi0mass} and \ref{fig:coplane}, the sPlot method is expected to remove all the background. Any error in the subtraction of the background by the weights-based method is expected to be less than the effect of not subtracting the background. Hence, the difference between $\Sigma$ extracted from the cuts and sPlot subtracted results is used as a conservative estimate of this systematic uncertainty~\cite{Mullen:2020fgd}. Values for this systematic uncertainty were calculated for each $E_{\gamma}$ and $\theta$ bin with a mean value of $3\%$. 
For the systematic uncertainty in the degree of linear polarization there are two factors, first the uncertainty in the SAID solution for $\Sigma$ on $p\pi^0$, which was used to determine the polarization, and was estimated to be $2\%$. Second, the uncertainty on our extraction of the $p\pi^0$ asymmetries which had a main contribution from the background subtraction which was estimated in a similar manner to the $n\pi^0$ background subtraction and found to be $4\%$. Adding these two factors in quadrature gives an overall $4.5\%$ systematic uncertainty in our $\Sigma$ results due to the linear polarization. 
Other uncertainties were found to be much smaller than these sources: acceptance effects, such as the neutron detection efficiency, cancel due to the polarization flip; polarization degree and luminosity asymmetries were incorporated into the likelihood fit; and unbinned fits were used, removing binning artifacts from the results.

\section{Multipole Analysis}
\label{Sec:PWA}

The SAID~\cite{Workman:2012jf}, MAID~\cite{Drechsel:2007if}, and Bonn-Gatchina~\cite{Anisovich:2009zy} analyses use different fit formalisms to extract the partial-wave (multipole) amplitudes underlying different data-sets. Comparing the different resulting amplitudes gives an estimate of the systematic errors inherent in the process.

For the MAID analysis, which was completed in 2007, the most recent data-sets have not been included and this must be considered when making comparisons \cite{Drechsel:2007if}. The MAID unitary isobar approach applies a Breit-Wigner resonance plus background model, guaranteeing unitarity up to the two-pion production threshold.

The Bonn-Gatchina method fits a wider range of reactions utilizing elements of the K-matrix and P-vector approaches~\cite{Anisovich:2009zy}. Both pion and photo-induced reactions are included in a multi-channel fit. Reactions with three-body final states are included using an event-based likelihood fit. The elastic pion-nucleon reaction is fitted based on existing amplitudes. The various data types are fitted with the possibility of renormalization and weighting.

The SAID method is an extension of the Chew-Mandelstam K-matrix approach used to fit pion-nucleon elastic scattering and $\eta N$ production data. The resonance spectrum is fixed from this fit~\cite{Arndt:2006bf} and only the photo-couplings are allowed to vary. This differs from the MAID and Bonn-Gatchina analyses, which can add new resonances to improve the agreement with data. The formalism has built-in cuts associated with the $\pi\Delta$, $\rho N$, and $\eta N$ thresholds but only single-pion photoproduction data are fitted. Data have been weighted and renormalized in previous fits. No weighting and only renormalization at the one percent level was utilized in fitting the present set of $\Sigma$ data.

For each angular distribution, a normalization constant ($X$) and its uncertainty ($\epsilon_X$) were assigned. The quantity $\epsilon_X$ is generally associated with the normalization uncertainty (if known). The modified $\chi^2$ function to be minimized is given by
\begin{equation}
    \chi^2 = \sum_i\left(\frac{X\eta_i - \eta_i^{exp}}{\epsilon_i}\right)^2 + \left(\frac{X - 1}{\epsilon_X}\right)^2 ,
    \label{eq:eq7}
\end{equation}
where the subscript $i$ labels the data points within the distribution, $\eta_i^{exp}$ is an individual measurement, $\eta_i$ is the corresponding calculated value, and $\epsilon_i$ represents the angular-dependent statistical uncertainty. The total $\chi^2$ is then found by summing over all measurements. This re-normalization freedom is often important in obtaining the best SAID fit results. For other data analyzed in the fit, such as the total cross sections and excitation data, the statistical and systematic uncertainties were combined in quadrature and no re-normalization was allowed.

In fitting the present set of $\Sigma$ data, an overall angle-independent systematic uncertainty of $1\%$ was used for $\epsilon_X$ in Eq.~(\ref{eq:eq7}). The resulting values for $X$ remained within $\epsilon_X$ of unity on average.

\begin{figure}[htb!]
\vspace{-10mm}
\centerline{\includegraphics[height=0.45\textwidth, angle=90]{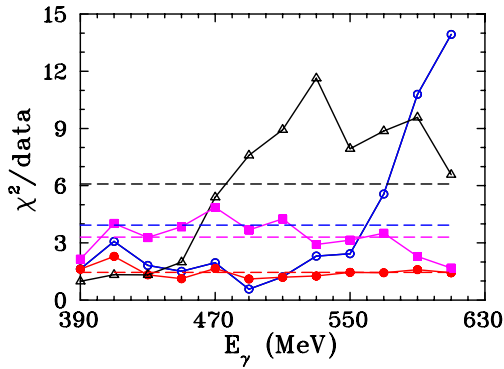}}

\vspace{-5mm}
        \protect\caption{Comparison of $\chi^2$ per data point for the previous SAID solution MA19~\protect\cite{Briscoe:2019cyo} and MAID2007~\protect\cite{Drechsel:2007if} 
        applied to the present A2 data with blue open circles and black open triangles, respectively. The new SAID solution MU22 (MUXX) with red full circles and 
        magenta full squares obtained after adding the present A2 data (MUXX does not 
        have any $\gamma n\to\pi^0n$ data). Also shown are the fit $\chi^2$ per data point 
        values averaged over each energy bin $E_\gamma$, where the horizontal dashed 
        lines are for the MU22 (red), MUXX (magenta), MA19 (blue), and MAID2007 (black) 
        solutions. The solid lines connecting the points are included only to guide 
        the eye.} 
        \label{fig:chi}
\end{figure}

A revised SAID multipole analysis has been completed, including the present set of $\overrightarrow{\gamma} n\to\pi^0 n$ $\Sigma$ data. This new global energy-dependent solution has been labeled as MU22. The overall fit quality of the present MU22 and previous MA19~\cite{Briscoe:2019cyo} SAID fits is compared with the MAID2007~\cite{Drechsel:2007if} solution in Tables~\ref{tab:tbl2} and \ref{tab:tbl3}. The inclusion of the present A2 data set provides a fit with significantly improved $\chi^2/data$, specifically at higher energies, in comparisons between the $\pi^0n$ fits and data ($\chi^2/data$ for MA19 = 3.93 and $\chi^2/data$ for MU22 = 1.44) as shown in Fig.~\ref{fig:chi} and Table~\ref{tab:tbl2}. This demonstrates the influence of these asymmetry measurements with their small uncertainties. The overall comparison of the MA19 and MU22 solutions shows that the fit $\chi^2/data$ values are essentially unchanged for $\pi^0p$ and $\pi^+n$ channels. The $\chi^2$ per data point including all available data and the present A2 data for MA19 and MU22 (with MAID2007) is given in Table~\ref{tab:tbl3}. 
\begin{table}[htb!]

\centering \protect\caption{$\chi^{2}$ per data point for new A2 data. 
        Predictions are from MAID2007~\protect\cite{Drechsel:2007if} and the 
        SAID fit MUXX (no world $\pi^0n$ data fitted), an older SAID fit to 
        existing data (MA19~\protect\cite{Briscoe:2019cyo}), and the SAID fit 
        including present data (MU22).}
\vspace{2mm}
{%
\begin{tabular}{|cc|}
\hline
Solution & $\chi^2/(\pi^0n$ data) \tabularnewline
\hline
MU22     &  275/189=1.46       \tabularnewline
MUXX     &  624/189=3.30       \tabularnewline
MA19     &  743/189=3.93       \tabularnewline
MAID2007 & 1151/189=6.09       \tabularnewline
\hline
\end{tabular}} \label{tab:tbl2}
\end{table}

Additionally, an alternative MUXX solution was generated excluding all world $\gamma n\to\pi^0n$ data and show results in Fig.~\ref{fig:610} and Table~\ref{tab:tbl2}. The excellent comparison of the isospin-predicted $\Sigma$ to the data strongly suggests the systematics in the new data are well under control. The comparisons with MAID are interesting as both solutions use isospin symmetry to predict the $\pi^0n$ observables based on the available data from the other three charge channels in 2007 and 2020. 
\begin{table}[htb!]

\centering \protect\caption{$\chi^{2}$ per data point values for all charge 
        channels covering the energy range from 155~MeV to 1000~MeV. Fits as 
        described in Table~\protect\ref{tab:tbl2}.}
\vspace{2mm}
{%
\begin{tabular}{|ccc|}
\hline
Solution & $\chi^2/(\pi^0p$ data) & $\chi^2/(\pi^+n$ data) \tabularnewline
\hline
MU22     &13274/9534=1.39& 7454/4039=1.85\tabularnewline
MUXX     &13171/9534=1.38& 7259/4039=1.80\tabularnewline
MA19     &12565/9534=1.32& 7461/4039=1.85\tabularnewline
MAID2007 &73638/9534=7.72&14599/4039=3.61\tabularnewline
\hline
Solution & $\chi^2/(\pi^0n$ data) & $\chi^2/(\pi^-p$ data) \tabularnewline
\hline
MU22     & 2345/ 798=2.94& 5879/3456=1.70\tabularnewline
MUXX     & 7639/ 798=9.57& 5384/3456=1.56\tabularnewline
MA19     & 2649/ 798=3.32& 5999/3456=1.74\tabularnewline
MAID2007 & 4846/ 798=6.07&15365/3456=4.45\tabularnewline
\hline
\end{tabular}} \label{tab:tbl3}
\end{table}

\section{Results and Interpretation}
\label{Sec:Results}

A comprehensive set of $\Sigma$ data for $\overrightarrow{\gamma} d\to\pi^0n(p)$ at 12 photon energies has been determined with the CB and TAPS spectrometers using a tagged photon beam at incident photon energies from 390 to 610~MeV. The present $\Sigma$ data cover the resonance region from above the maximum of the $\Delta$-isobar to the Roper resonance. 
\begin{figure}[htb!]
\centerline{\includegraphics[height=0.27\textwidth]{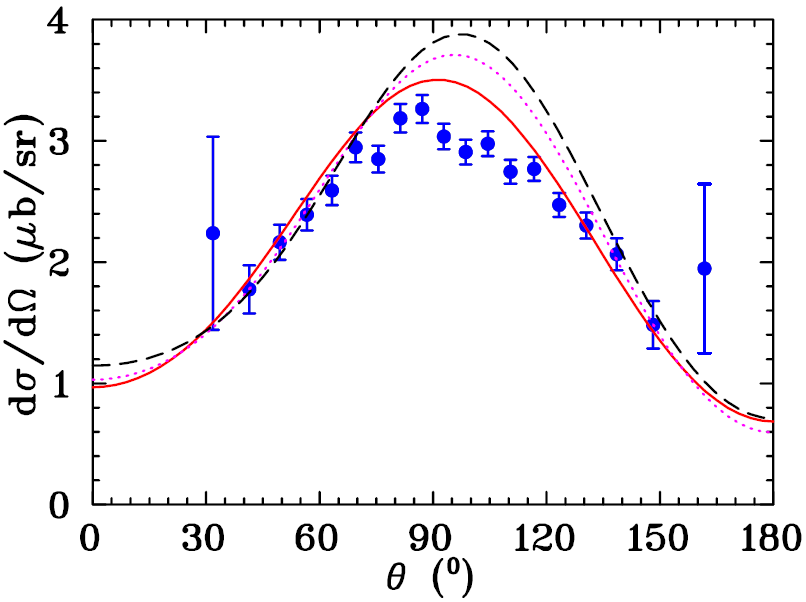}}
\centerline{\includegraphics[height=0.27\textwidth]{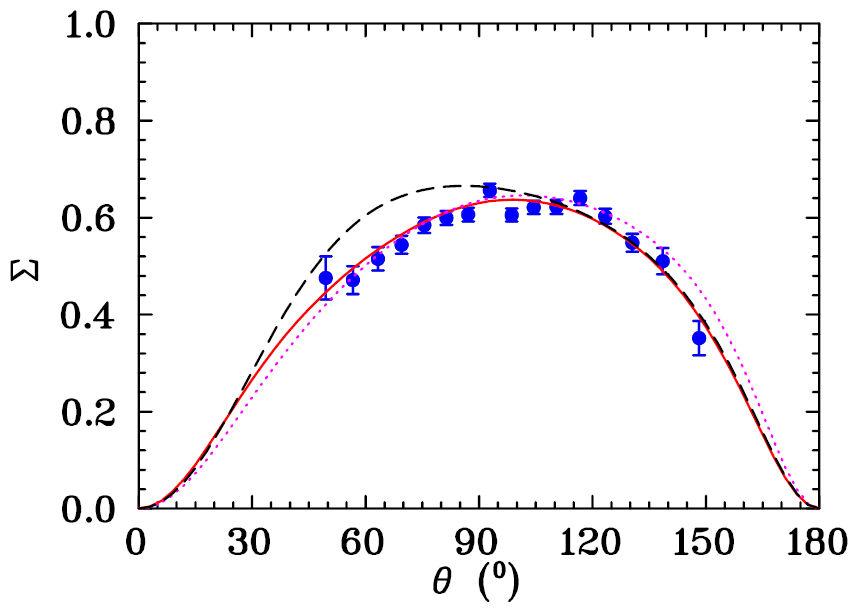}}
\centerline{\includegraphics[height=0.27\textwidth]{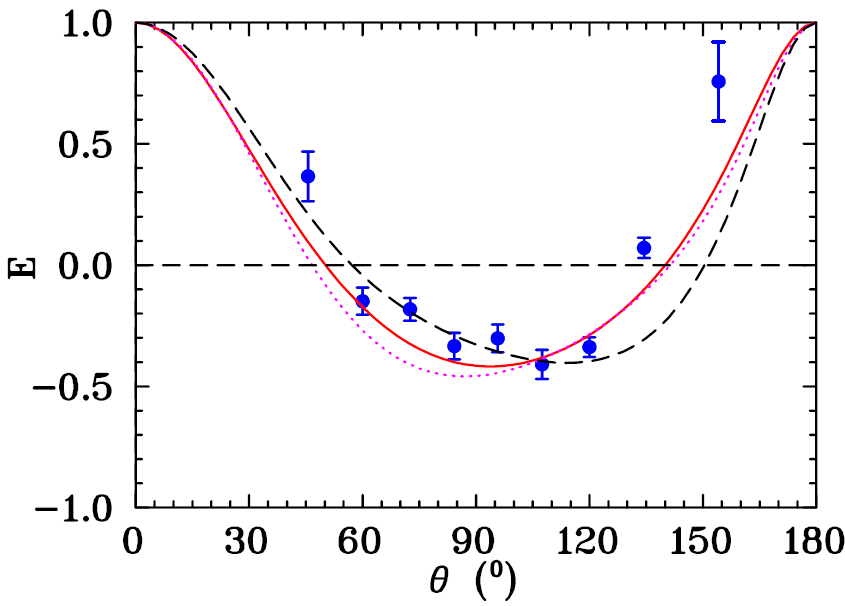}}

\protect\caption{$d\sigma/d\Omega$, $\Sigma$, and $E$ observables for 
        $\overrightarrow{\gamma} n\to\pi^0n$ (blue full circles). 
        All data were produced by the A2 Collaboration at MAMI:
        $d\sigma/d\Omega$ at $E_\gamma = 610$~MeV (top) are from 
        Ref.~\protect\cite{Briscoe:2019cyo}, $\Sigma$ at $E_\gamma = 
        610$~MeV (middle) are present measurement, and  $E$ at 
        $E_\gamma = 603$~MeV (bottom) are from 
        Ref.~\protect\cite{Dieterle:2017myg}.
        New SAID solutions MU22 (MUXX) are shown by red solid (magenta 
        dotted) curves and MAID2007~\protect\cite{Drechsel:2007if} 
        by black dashed curves. 
        \label{fig:610}}
\end{figure}

The SAID MA19~\cite{Briscoe:2019cyo}, Bonn-Gatchina BG2014-02~\cite{Gutz:2014wit}, and MAID2007~\cite{Drechsel:2007if} curves shown in Fig.~\ref{fig:sigma} did not include the present A2 data in their fits. In addition, the MAID2007 fit does not include measurements after 2007. MU22 includes all previous measurement and includes the A2 data. All fits and predictions agree well for the lowest energy where the $\Delta$ resonance dominates. The angular distribution retains the shape of SAID MA19 up until the highest energies where some larger deviations become apparent. 

\begin{figure*}[htb!]
\centerline{\includegraphics[height=0.4\textwidth,angle=90]{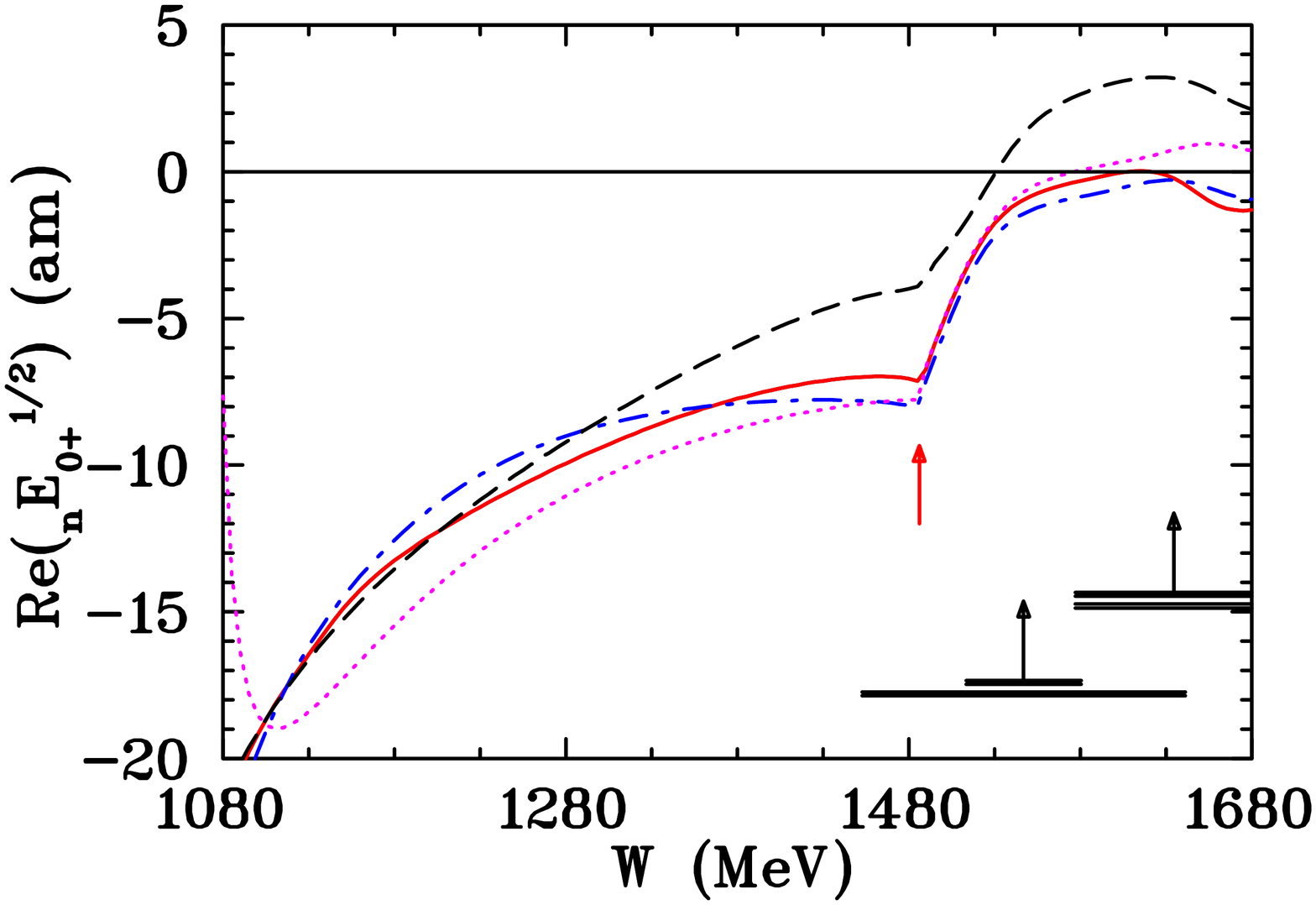}
            \includegraphics[height=0.4\textwidth,angle=90]{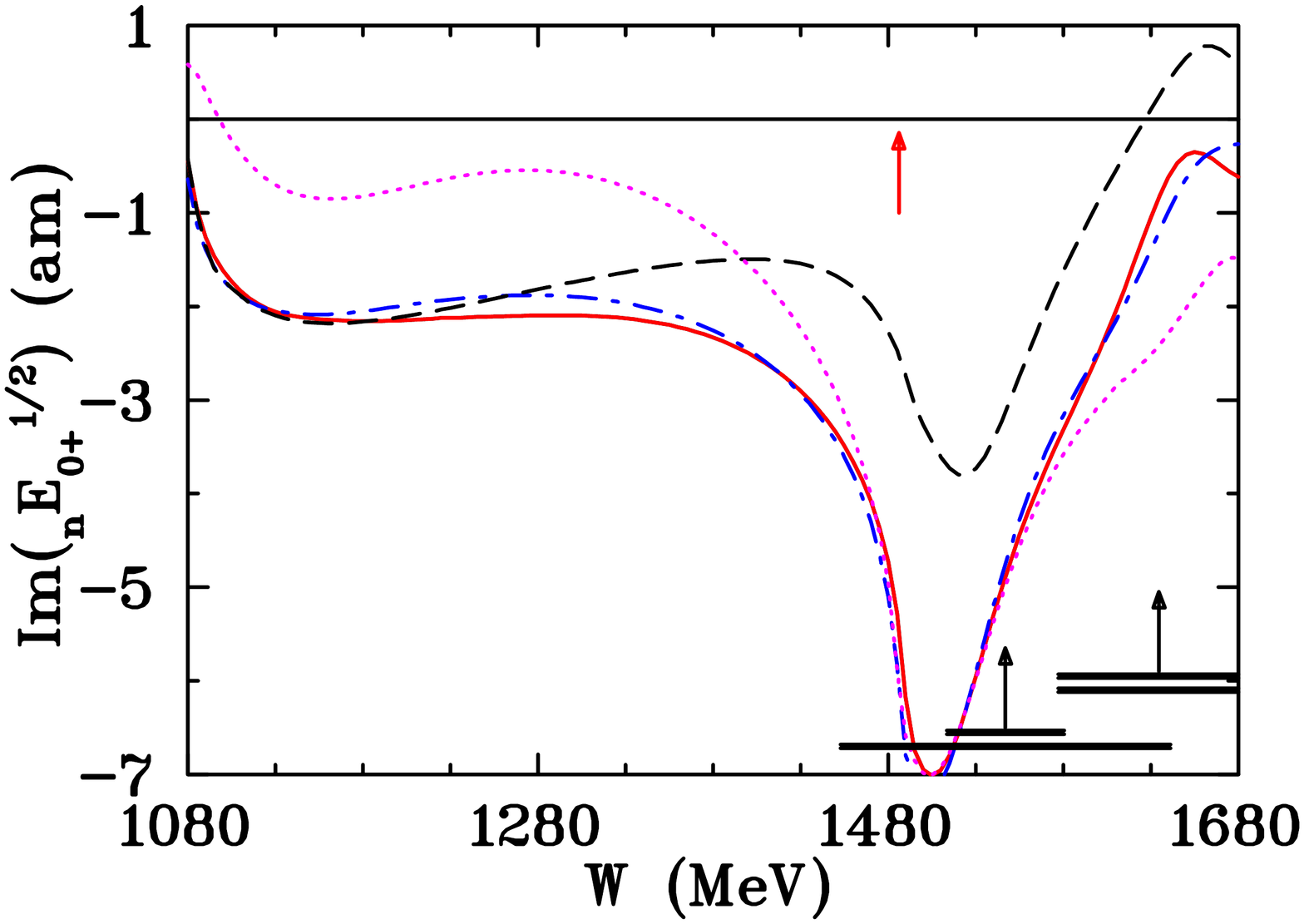}}
\centerline{\includegraphics[height=0.4\textwidth,angle=90]{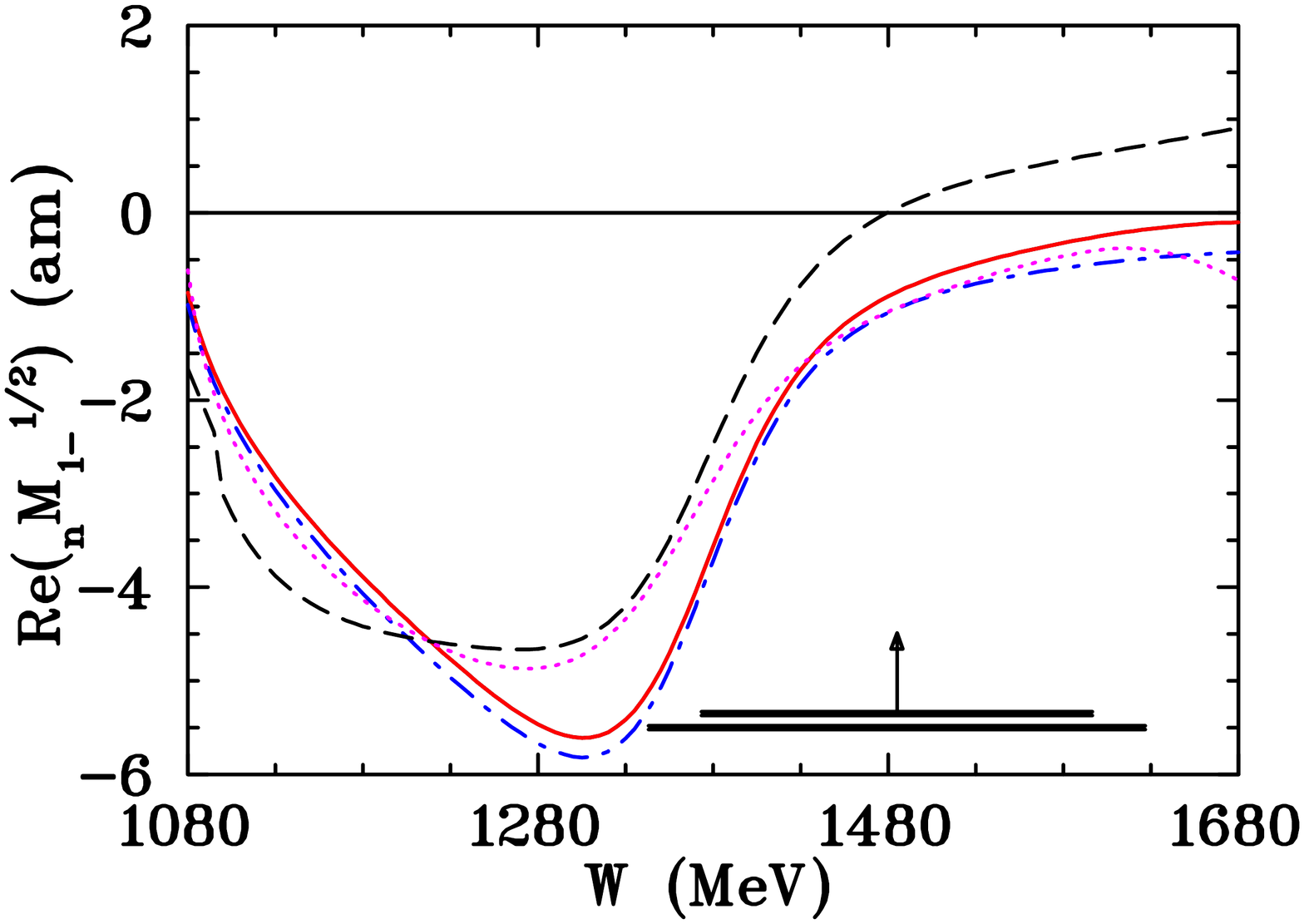}
            \includegraphics[height=0.4\textwidth,angle=90]{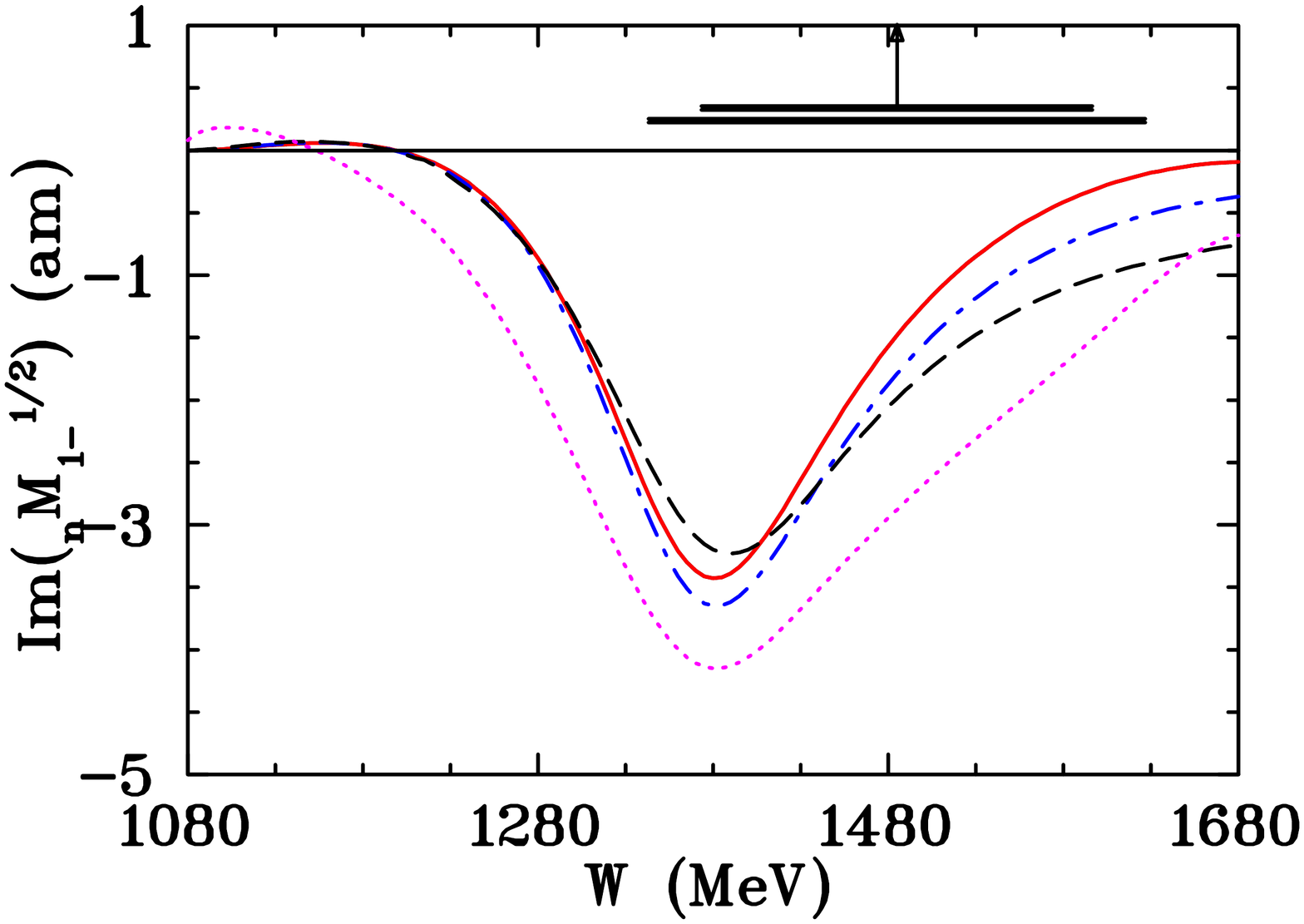}}
\centerline{\includegraphics[height=0.4\textwidth,angle=90]{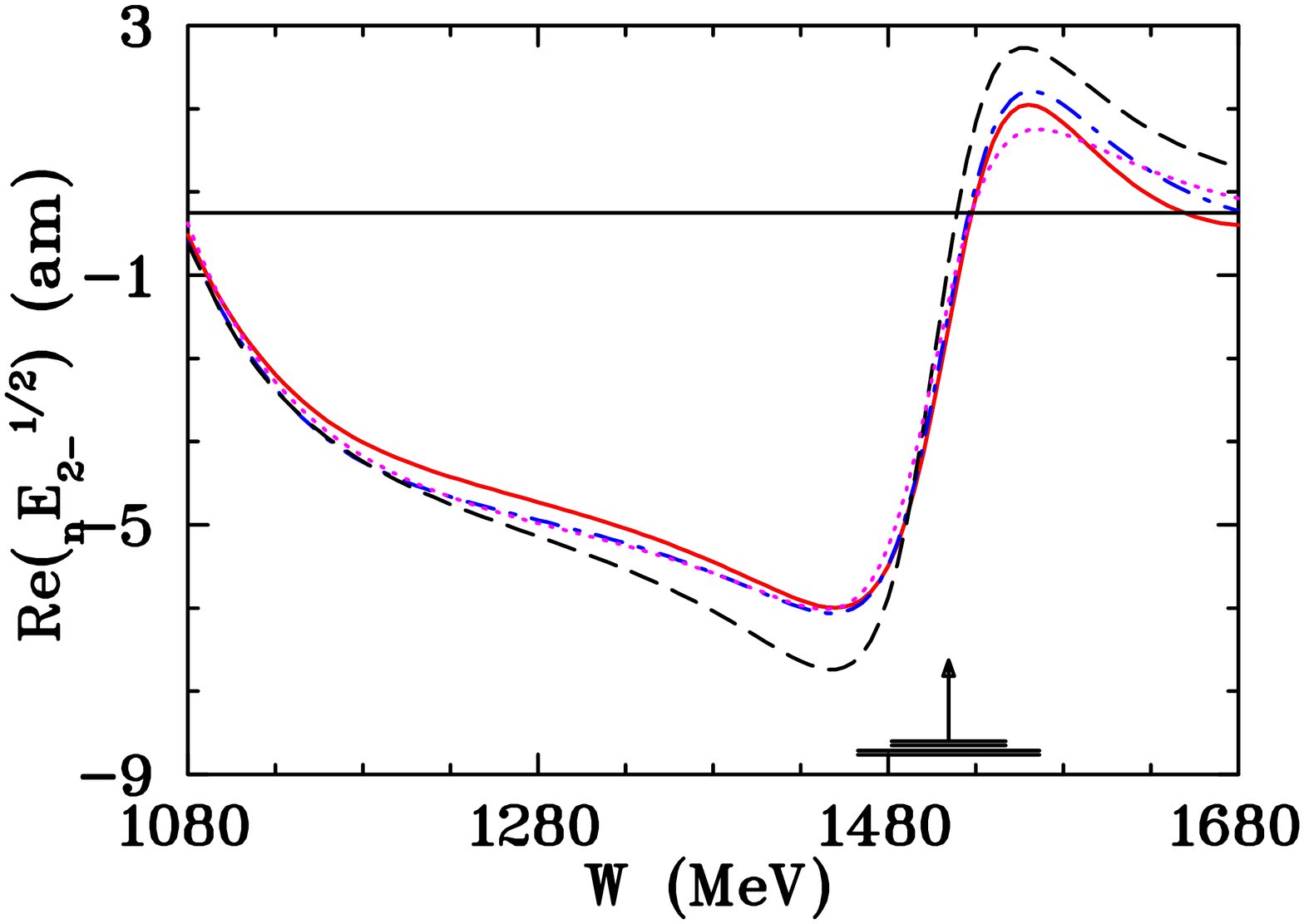}
            \includegraphics[height=0.4\textwidth,angle=90]{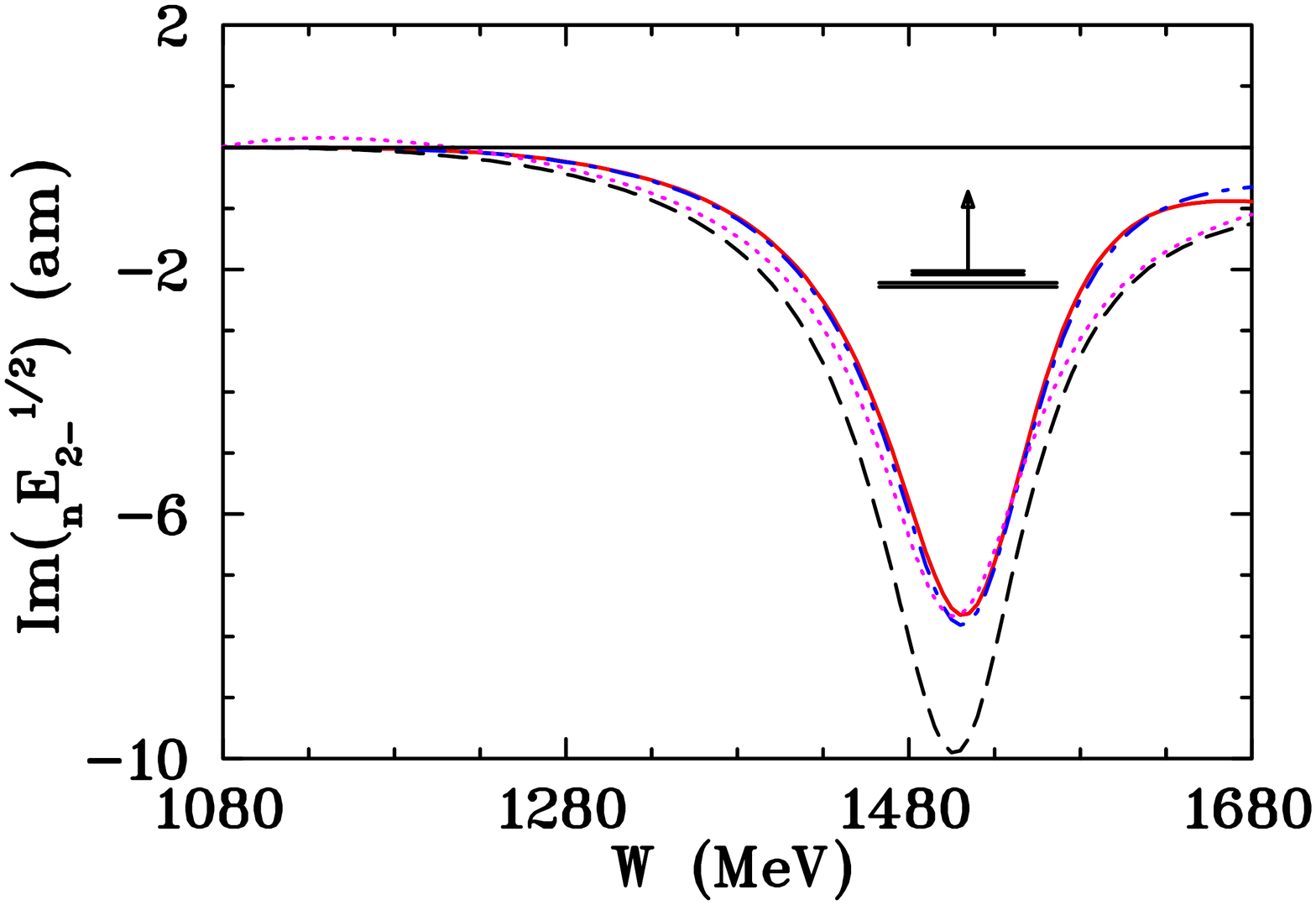}}
\centerline{\includegraphics[height=0.4\textwidth,angle=90]{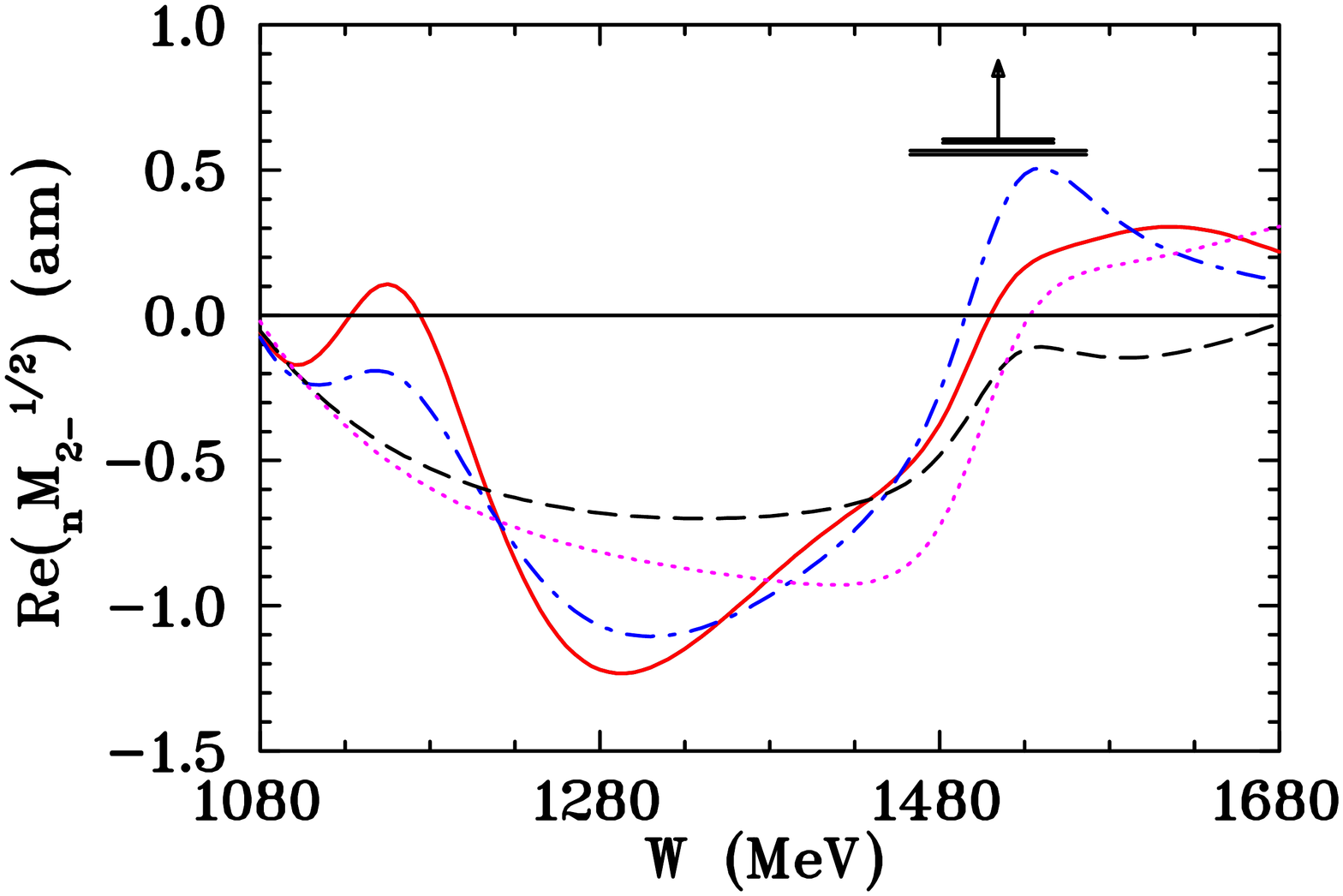}
            \includegraphics[height=0.4\textwidth,angle=90]{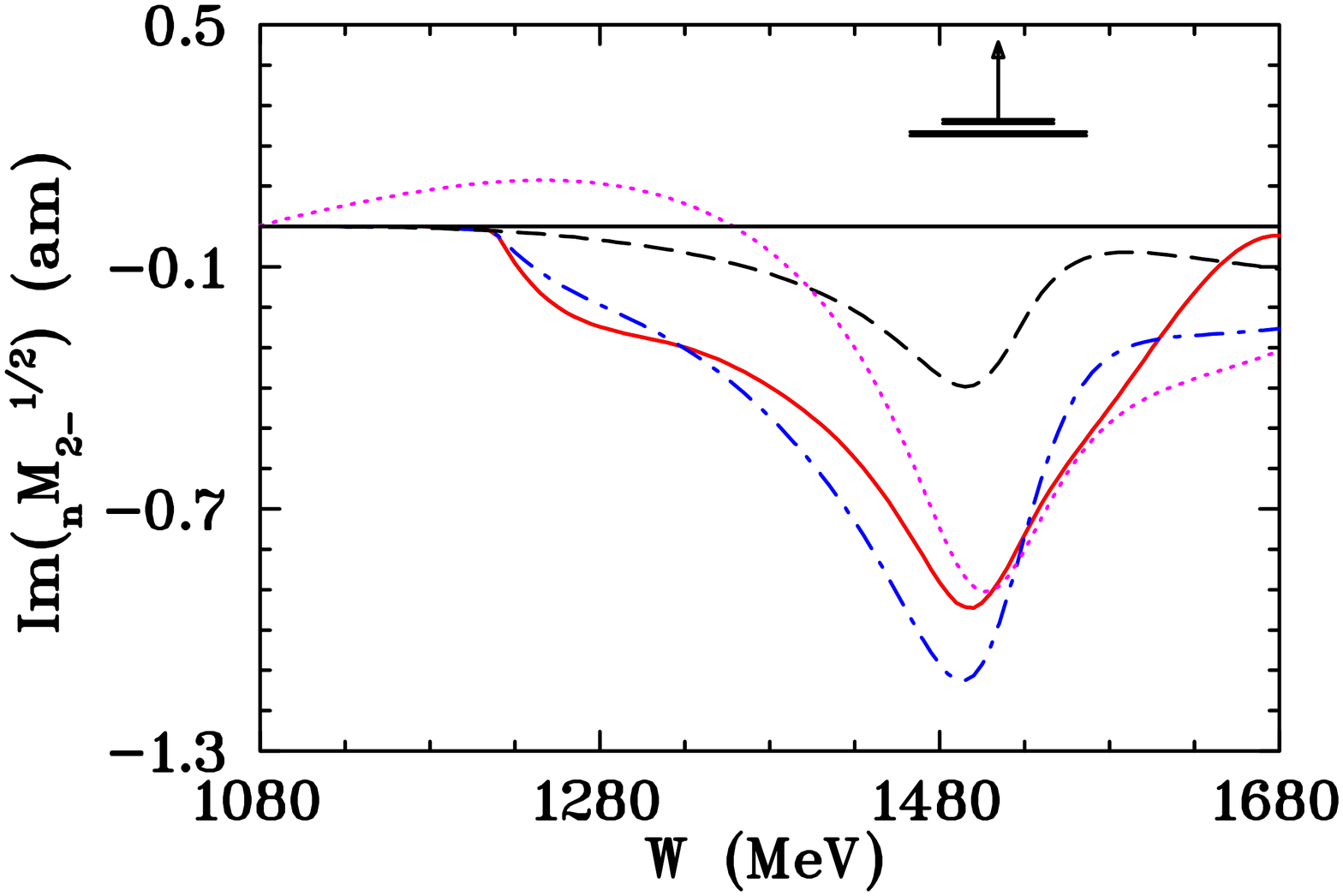}}

\protect\caption{Selected neutron multipole $I$ = 1/2 amplitude from 
        threshold to $W$ = 1.68~GeV ($E_\gamma$ = 1030~MeV) for the 
        states $0^+$, $1^-$, and $2^-$. 
        The blue dash-dotted (red solid) curves correspond to the SAID 
        MA19~\protect\cite{Briscoe:2019cyo} (new MU22 including present 
        A2 data) solution. The magenta dotted (black dashed) curves give 
        the Bonn-Gatchina BG2014-02~\protect\cite{Gutz:2014wit} 
        (MAID2007~\protect\cite{Drechsel:2007if}). 
        The vertical black arrows indicate Breit-Wigner (BW) mass ($W_R$), and 
        horizontal bars show full ($\Gamma$) and partial ($\Gamma_{\pi N}$) 
        widths of resonances extracted by the BW fit of the $\pi N$ data 
        associated with the SAID solution SP06~\protect\cite{Arndt:2006bf}.
        Vertical red arrows show the $\eta$ meson production threshold.
        \label{fig:ampl1}}
\end{figure*}

Exploring the effectiveness of isospin symmetry to predict $\pi^0n$ observables, in Fig.~\ref{fig:610} the fit (MU22) is compared to predictions from MUXX and MAID2007 for $d\sigma/d\Omega$, $\Sigma$, and the double-polarization asymmetry $E$. At this energy, the qualitative features are generally reproduced, particularly for the current data. In the fit, cross sections have larger angle-independent systematic uncertainties and the renormalization factor from Eq.~(\ref{eq:eq7}) improves the description, but is not included in the plot. While the new data cover parts of the $\Delta$ and Roper resonance regions, selected isospin multipoles are compared up to a photon energy of 1~GeV. The isospin $3/2$ multipoles are taken as determined by the much larger proton-target database and only the isospin $1/2$ neutron multipoles are shown in Fig.~\ref{fig:ampl1}. Comparing MA19 to MA22, there are no significant changes seen in the $M_{1-}^{1/2}$ multipole. In addition, comparing imaginary parts of multipoles connected to the nearby $N(1535)$ ($E_{0+}^{1/2}$) and $N(1520)$ ($E_{2-}^{1/2}$ and $M_{2-}^{1/2}$), there is also good agreement between the SAID and Bonn-Gatchina plots. The resonance couplings for the $N(1535)$, $N(1440)$, and $N(1520)$ are expected to be in agreement with those reported in Ref.~\cite{Briscoe:2019cyo}.

As a final comment on the predictive ability of our fit, excluding all $\pi^0n$ data, a comparison of Tables~\ref{tab:tbl2} and \ref{tab:tbl3} shows that fit MUXX is much less successful between the upper energy limit of the present experiment and 1~GeV in the photon energy. This change is due mainly to poor compatibility with GRAAL $\Sigma$ data~\cite{DiSalvo:2009zz}. 

Our results for $\Sigma$ for $\overrightarrow{\gamma} d\to\pi^0n(p)$ consist of 189 experimental points and are available from the SAID database~\cite{Briscoe:2020zzz}, where systematic uncertainties for each bin have been added in quadrature.

\begin{acknowledgements}
This work was supported in part by the UK Science and Technology Facilities Council (STFC Grants No. 57071/1 and No. 50727/1), the U.~S.~Department of Energy, Office of Science, Office of Nuclear Physics, under Awards No. DE–FG02–01ER41194, No. DE–-SC0016583, No. DE-–SC0016582, and No. DE-–SC0014323. This work was supported by Schweizerischer Nationalfonds (Grants No. 200020--132799, No.~121781, No.~117601, and No.~113511), \\ Deutsche Forschungsgemeinschaft (SFB Grant No.~443, \\ No. SFB/TR~16, No. SFB~1044), DFG--RFBR (Grant No. 05–-02-–04014),  European Community Research Infrastructure Activity (FP6), the U.~S.~DOE, U.~S.~NSF, and NSERC (Grant No. SAPPJ–-2018-–00020) Canada.
This publication is part of a project that has received
funding from the European Union’s Horizon 2020 research and innovation programme under grant agreement
STRONG – 2020 - No 824093.
We would like to thank all the technical and nontechnical staff of MAMI for their support.
\end{acknowledgements}


\end{document}